# 9    Model Evolution and Management

Tihamer Levendovszky[1], Bernhard Rumpe[2],
Bernhard Schätz[3], and Jonathan Sprinkle[4]

[1] Vanderbilt University, Nashville, TN, USA
`tihamer@isis.vanderbilt.edu`
[2] RWTH Aachen University, Germany
`http://www.se-rwth.de`
[3] fortiss GmbH, München, Germany
`schaetz@fortiss.org`
[4] University of Arizona, Tucson, AZ, USA
`sprinkle@ECE.Arizona.Edu`

**Abstract.** As complex software and systems development projects need models as an important planning, structuring and development technique, models now face issues resolved for software earlier: models need to be versioned, differences captured, syntactic and semantic correctness checked as early as possible, documented, presented in easily accessible forms, etc. Quality management needs to be established for models as well as their relationship to other models, to code and to requirement documents precisely clarified and tracked. Business and product requirements, product technologies as well as development tools evolve. This also means we need evolutionary technologies both for models within a language and if the language evolves also for an upgrade of the models.

This chapter discusses the state of the art in model management and evolution and sketches what is still necessary for models to become as usable and used as software.

## 9.1    Why Models Evolve and Need to Be Managed?

### 9.1.1    Introduction

Any complex set of artifacts needs to be managed, and models are certainly no exception—especially given that models are used to help manage complexity, and are therefore used in complex projects. Even though models do reduce the project's complexity, projects often have a complexity that even clever abstractions cannot transcend; thus, the models used in such a development project either become complex themselves or there are very many (perhaps, heterogeneous) models—or both.

This complexity also means that we cannot just create and forget models, but we must continuously evolve a model when adding new information, after quality reviews, redesigns against flaws and (in particular) according to changing requirements.

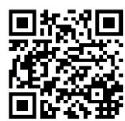

### 9.1.2   Model Management

*Model management* is the coordination between model-driven engineering (MDE) artifacts and resources, such as models, metamodels, transformations, correspondence, versioning, etc. [1]. Thinking in terms of these global (and entirely model-based) solutions has also been referred to as "megamodelling" [2].

Model management helps us to understand the status of models during the development and the maintenance process as well as how models relate to each other and to other artifacts of a development project. Proper model management is therefore a basic necessity to run a model-based development process (of a certain complexity). Most of the model management techniques are primarily for the developers themselves to simplify their life, increase efficiency when assessing or evolving models and ensure that less tedious work has to be done or even redone when requirements or technical components evolve. Other techniques in particular high-level reports are for the project management— to understand and measure progress and project status.

Among model management techniques, we distinguish the following main categories:

- Checking for structural qualities of models, such as consistency and completeness on the one hand, but also guideline checking for additional quality attributes like readability or evolvability
- Transforming of models, including constructive and declarative descriptions of structural relations between models
- Versioning and version control of models, including the reporting of differences between model versions and merging of independently changed models with a common ancestor (so called "three way merge")

While we consider code generation, analysis and simulation tools the part of model management, we will not concentrate on these issues, but rather on the issues that treat the models as artifacts. Code generation, model-based analysis and simulation, etc., are undergoing heavy discussion and development right now, and there is a tremendous variety of approaches, from interactive one-shot generation of code-frames to repeatable fully automatic generation and customization/configuration of complete and deployable software components.

Model management for embedded systems includes additional twists on managing models for non-software artifacts. For example in product lines or other kinds of evolvable systems it is necessary to keep track of the connections between software and the mechanical, hydraulic or electric parts of an engine and their controlling software. This needs integrated models and thus integrated model management.

Model management for embedded systems includes additional twists on managing models for non-software artifacts. For example in product lines or other kinds of evolvable systems it is necessary to keep track of the connections between software and the mechanical, hydraulic or electric parts of an engine and their controlling software. This needs integrated models and thus integrated model management.

### 9.1.3   Model Evolution

According to our taxonomy, model evolution is one part of the more general model management issue. However, model evolution has many variations, and it is an important piece of model management. That is why we concentrate on model evolution, both from methodical as well as from a technical viewpoint in Section 9.3, because many of its concepts can be generalized to reflect versioning, interchange, tracking, consistency, etc.

In this paper, we use the term *model evolution* to refer to techniques to adapt existing models, as well as their related context, according to evolving needs. This context includes other models, code, tests, informal descriptions etc. that all might be affected when a model's content is changed. Evolution occurs when requirements or technology change as well as when improvements are made.

Model evolution may be necessary because of quality deficiencies according to two categories: if the model does not fit its context anymore; or if the representation of a model is bad and needs to be enhanced (e.g., to increase readability). As an important new problem, we also see the need of models to evolve together with the underlying language in which the model is expressed. As domain-specific languages (DSLs) [3] increase in popularity, and are often developed within or in parallel to the project, the evolution of a language quite often enforces the evolution of the models as well.

In this paper, we use *language evolution* to refer to techniques to evolve a modelling language according to domain or technology change, including the parallel evolution of that language's models and tools.

### 9.1.4   Chapter Outline

The rest of this chapter is organized as follows. Section 9.2 discusses the above mentioned techniques of model management. Section 9.3 deals with management of models, both from methodical as well as from a technical viewpoint, and in Section 9.4, we examine a particular instance of evolution driven by evolution of a domain-specific language. Furthermore, we examine in detail the case in which the evolution happens in small steps rather than abrupt changes.

## 9.2   Model Management

With a more widespread use of industrial-scale models throughout the development process, 'classical' problems found in code-oriented development start to impact a model-based development in a similar manner: The legacy of long-living models requires to address issues such as modeling standards and the quality of models, or model-versioning and -merging.

While for a code-based development many solutions have already been put in place to counteract these problems, in mode-based development these solutions are increasingly becoming available. For many issues, e.g., conformance or consistency analysis, the use of concept-rich, domain-specific models with a precise interpretation even allow to improve existing solutions for a code-based approach with its weaker-structured, more generic form of representation. For other issues, e.g., the merging or versioning, the additional complexity introduced by the rich structure of models leads to new challenges.

### 9.2.1   Model Quality and Modeling Standards

The application of quality constraints on the construction of software products – generally in the form of standards providing rules and guidelines for the construction of code – has repeatedly shown its merits in the development process concerning the quality of the product, both with respect to reliability as well as the maintainability. While the use of models in a development process provides an important constructive form of quality assurance, the rich and rigid structure and the domain-specific nature of the models used in the development allow to add new forms of quality constraints to the development process.

Therefore, quality constraints have increasingly gained attentions. Modeling tools provide mechanisms to ensure that the model under development respect modeling guidelines (e.g., the MAAB [4] guidelines for the Matlab/Simulink tool such [5]). Depending on the extent of these guidelines, they provide some constraints on the presentation (e.g., start state in top left area), the structure (e.g., number of interface elements), or even their interpretation (e.g., no lacking transitions). These constraints help to improve quality aspects like understandability, maintainability or even correctness.

Since obviously, not every syntactically correct model is a good model, there are many additional constraints that need to hold for a model to make it readable, changeable, or usable, etc. Depending on the nature of the constraints – and subsequently their implementation in a corresponding modeling framework – different kinds of conditions can be classified:

**Syntactic constraints** are immutable constraints enforced by construction on the structure of the model, ensuring that a model conforms to its metamodel and thus can be processed, stored and loaded, etc. Such a constraint e.g. ensures that a transition always is connected to a start and an end state.

**Well-formedness constraints** are constraints on the structure of the model enforced at specific steps in the development process, to ensure that the model is structurally sound. Such a constraint, e.g., ensures that used variables are actually defined and have the appropriate type. Such conditions are generally mechanically checked, e.g., upon processing or editing models.

**Semantic constraints** are constraints on the interpretation of a model, to ensure that a model is semantically correct. Such constraint, e.g., ensures that a state-transition model is not non-deterministic or incomplete. These conditions cannot always be effectively checked mechanically, facing the typical problems of model-checking approaches.

Note that shifting constraints between the first and the second class influences the rigidness of the development process.

Since model-based development increasingly deals with 'mega-modeling' issues [6] like large-scale, distributed models including linking models from heterogeneous domains – or meta-models – a second taxonomy builds around relations within and between models and domain-languages:

**Intra-model conditions** are defined over a single given model and thus can be formalized within the same domain-language and checked on a single model directly. A typical example is the type-correctness of a single dataflow model.

**Inter-model conditions** are conditions define for a set of models from the same domain, which still can be formalized within the same language but are checked on a set of models. A typical example is the interface consistency between models describing the environment of a system and its internal structure.

**Inter-language conditions** are conditions defined on a set of model of different modeling domains or languages. These conditions do not only require to check several models, but also can only be expressed in a super-language for these different languages. A typical examples is the consistency between different abstractions or viewpoints of a system, e.g., in case of the UML with its various sub-languages the consistency between a sequence diagram of a component, its state machine, as well as its interface view.

As both taxonomies are independent of each other, each combination has its relevance in the practical application.

For the practical application of conformance constraints in the development process, support for the definition of constraints on the models of the product under development as well the automated enforcement of these constraints proves to be an important asset, improving and front-loading this form of quality assurance in a model-based approach.

By adding a mechanism to automatically check for the validity of these constraints with respect to a product model and report violations on the level of the product model, conformance analysis can be supported by translating constraints of the guidelines to conformance conditions and using this mechanism for their validation.

Since analyzing the conformance of a model of a product to a certain modeling guideline is becoming increasingly relevant for the model-based development process, corresponding functionalities are added to tools supporting this kind of process. For widely used tools such as MatLab [5], additional mechanisms – integrated into the tool itself or provided as a stand-alone checker – ensure that the model under development respect conformance constraints. However, since often those mechanisms use the API of the tools (e.g., *M-Script* for *MINT* or *Model Advisor* [5]; *Java* for *MATE* [7]; *MDL* for ConQAT [8]), conformance constraints are rather defined on the level of the concepts of implementation language than at level of the concepts of the application domain.

Here the fact can be exploited that the conceptual domain model allows to define a *conceptual product model* and furthermore *provides a vocabulary capable*

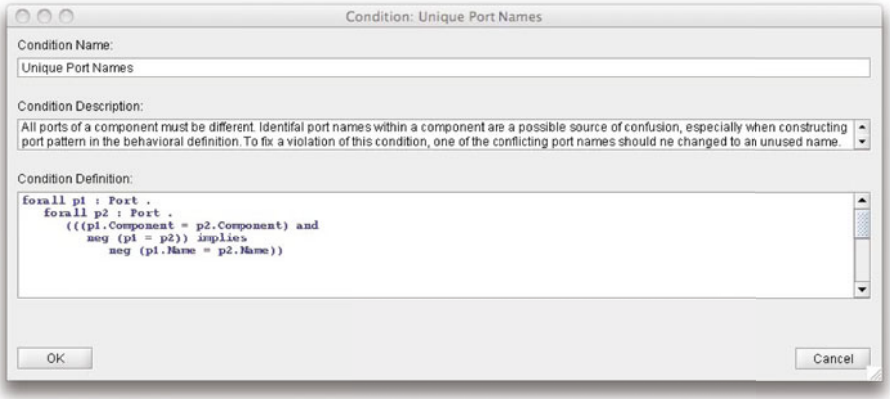

**Fig. 9.1.** Defining a Conformance Condition

*of defining properties* of such a product model. In order to enforce conformance constraints on products of a specific domain of application, this vocabulary must allow to *express these constraints as logical conditions* over the classes and associations of the concept model characterizing the domain of application. Obviously, this formalism should reflect the concept model as closely as possible, to abstract from the actual technical representation of the model of the product.

In [9], an approach is used for formalization of conformance constraints based on the OCL formalism [10]. [11] uses a simular formalism, based on predicate logic with the classes and associations of the concept model as first-class-citizens. Constraint checking can be performed by

– providing a checker separate from the tool for the construction of the product itself, generating a report as mentioned for the further improvement of the model; this technique is chosen in the former approach;
– integrating the checker into the construction tool, allowing direct navigation or direct application of improvement operations; this technique is chosen in the latter approach.

Independently of the degree of integration, this form of constraint analysis is performed in three steps:

(1) Definition of constraints, often combined with a grouping of similar constraints
(2) Checking of the validity of the constraints
(3) Inspection of the counterexamples for violated constraints.

In the following, these steps are demonstrated for the approach described in [11].

Although conformance analysis can be understood as checking the validity of conceptual conditions, for practical application, issues besides the evaluation have to be considered. Therefore, as shown in Figure 9.1, the definition of a consistency condition consists of a **constraint name** for selecting the the condition using the process interface, an informal **constraint description**, generally for the purpose of the condition and possible remedies in case the condition does not hold for the model, and an expression formalizing the condition definition.

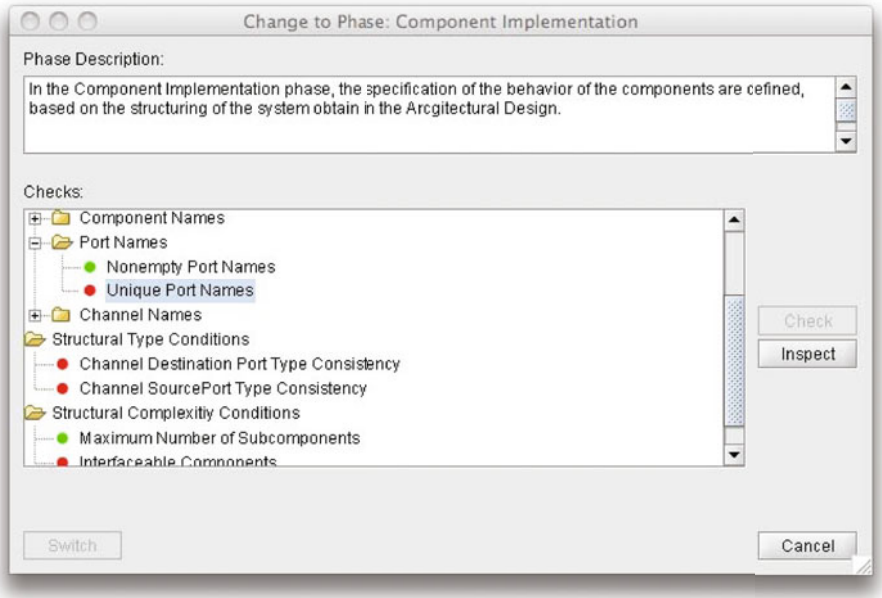

**Fig. 9.2.** Checked Conditions of a Phase

Besides the validity of the consistency condition, the collection of all model elements not satisfying the condition are an important result in case the condition is not satisfied. Accordingly, the user is rather interested in in all model elements satisfying the complement of the consistency conditions. To check the validity of the formalized constraints in the next step, the constraints are evaluated. If all conditions are satisfied, the validity is stated. Otherwise – as shown in Figure 9.2 – violated entry conditions like 'Unique Port Names' are reported, and can be inspected as described below.

When evaluating a constraint, besides returning its validity, the checker returns the set of all unsatisfying assignments of the quantified variables. This collection of model elements not satisfying the consistency condition is the set of all counterexamples throughout the product model violating the consistency constraint imposed on the model

As shown in Figure 9.3, this result is returned as a list of assignments; additionally, the informal description of the consistency check is presented to support the user in correcting the inconsistency. To simplify correcting a violated consistency condition, a simple navigation mechanism from such an assignment to the graphical representation of the model is provided: by activating an assignment from the list of inconsistencies, the corresponding editors containing the violating element is presented.

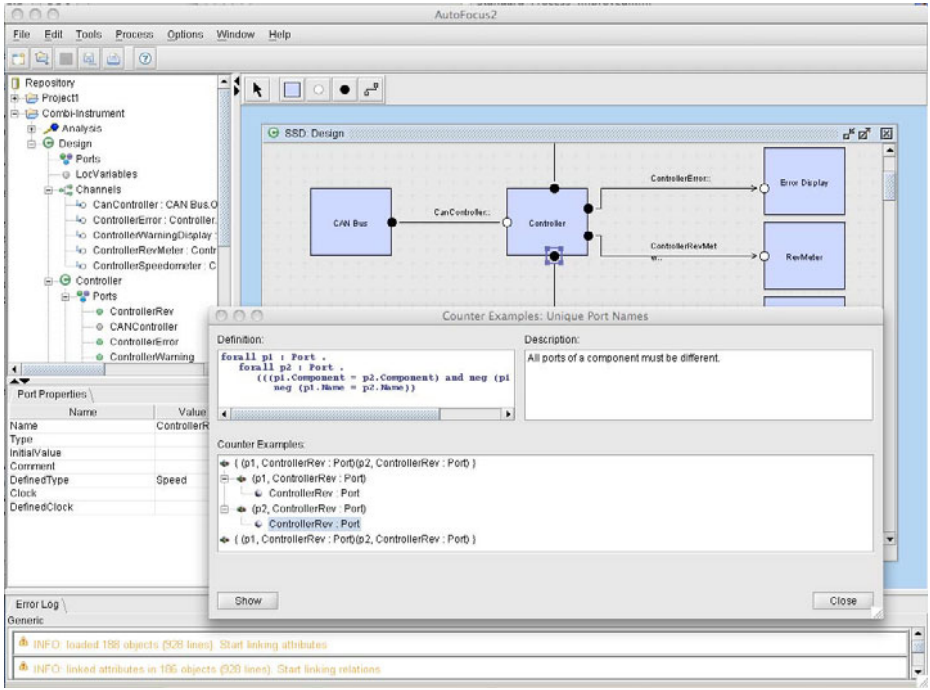

**Fig. 9.3.** Counterexamples of a Constraint Evaluation

Like the checking of modeling standards corresponding to the checking of coding standards, also other techniques found in a code-centric development are applicable for a model-based development process, e.g., the detection of model clones. In general, clones are *description fragments that are similar w.r.t. to some definition of similarity.* The employed notions of similarity are heavily influenced by the program representation on which clone detection is performed and the task for which it is used.

The central observation motivating clone detection research is that code clones normally implement a common concept. A change to this concept hence typically requires modification of all fragments that implement it, and therefore modifications of all clones, thus potentially increasing the maintenance effort. Additionally, clones increase description sizes and thus further increase maintenance efforts, since several maintenance-related activities are influenced by description size. Furthermore, bugs can be introduced, if not all impacted clones are changed consistently.

Here, detection of model clones like in [12] can improve the maintainability of evolving models, helping to identify redundant model fragments. Figure 9.4 shows the application of clone detection to dataflow languages as used in Simulink.

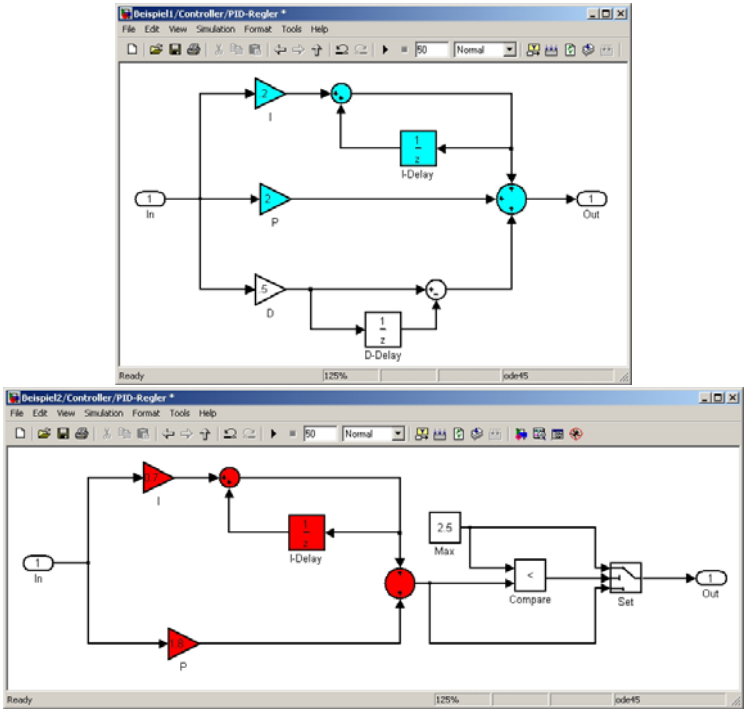

**Fig. 9.4.** Example for Clones in Dataflow Models

### 9.2.2  Model Transformation

Especially in a model-based approach with structure-rich system descriptions, automated development steps, – using mechanized transformations – have the potential to provide an important technique to improve the efficiency of the development process. Besides increasing efficiency, these structural transformations can offer consistency ensuring modification of models. There is a range of different applications for model transformation:

– Refactoring of models, e.g, to improve the architecture of a system, operating on a single model
– Merging of models, e.g., to consistently weave in standard behavior, operating on a set of model of the same language
– Translation of models, e.g., to generate a platform-specific model from a platform-independent model, operating on a set of models of different languages

However, for their effective application, frameworks providing these transformations should use formalisms to enable sufficiently abstract yet executable descriptions, support their modular definition by simple composition, and supply mechanisms for parameterization. Generally, these transformations are executed on the internal representation, called *abstract syntax*, of the models, but often defined on the representation of the model, called the *concrete syntax*.

Similarly to the case of the structural analysis of conceptual product models, the principle of transformation on the internal model representation makes use of the fact that, in a model-based approach, a product model comes with an *explicit representation of the abstract syntax* composed of domain-specific concepts and their associations; therefore, *transforming this structure* corresponds to transforming the product model.

By providing a *language capable of relating properties of those structures* of concepts and associations, a transformation can be understood as a *relation between a product before and after the transformation*. Then, by applying one argument of this relation to the model of the product under development and by providing a mechanism to *constructively compute* the other argument, the relation creates the transformed product. Thus, by formalizing standard operations of a development process as transformation relations, the process can be supported by mechanized operations. Examples for these operations are architectural refactorings of systems consisting of hierarchical components and connected via ports linked by channels; e.g., the pushing down of a component into a container component, making it a subcomponent of that container and requiring to split or merge channels crossing the boundary of the container component via the introduction or elimination of intermediate ports. Transformations like this structural refactoring or the semantically equivalent refactorings of state machines found in [13] are especially important since they leave the behavior of the modeled system unchanged.

Due to their generality, model transformations form the basis for many model-driven approaches ([14, 6, 15, 16]). In contrast to other development environment such as the Eclipse Refactoring plug-in, providing a fixed set of refactoring rules, a generic transformation mechanism allows the tool-user or tool-adaptor to enhance the functionality of the tool by defining domain-specific operations such as safe refactoring rules. Since models can be interpreted as graphs, transformation frameworks generally define operations as graph transformations, providing constructs to manipulate nodes (elements) and edges (relations) of a product model.

This generic approach is used in graph-based frameworks such as MOFLON/ TGG [17], VIATZRA [18], FuJaBa [19], DiaGen [20], AToM³ [21], or GME [22]. These approaches are based on graph-grammars or graphical, rule-based descriptions [23]. Basically, for the declaration of basic transformations the transformations are described in a pre-model/post-model style using graph-patterns. In triple-graph-grammar approaches [24] additionally a correspondence graph [25] is added to explicitly model mappings between (parts of) the pre- and post-model during transformation. Furthermore, those approaches often use extensions to enhance the patterns as well as to describe their compositions, such as OCL expressions, and state machines. Figure 9.5 shows the formalization of a rule for the merging of a channel in the push-down refactoring in the FuJaBa approach, using an extended object diagram notation with annotations to specify the creation or destruction of objects and links in a product model.

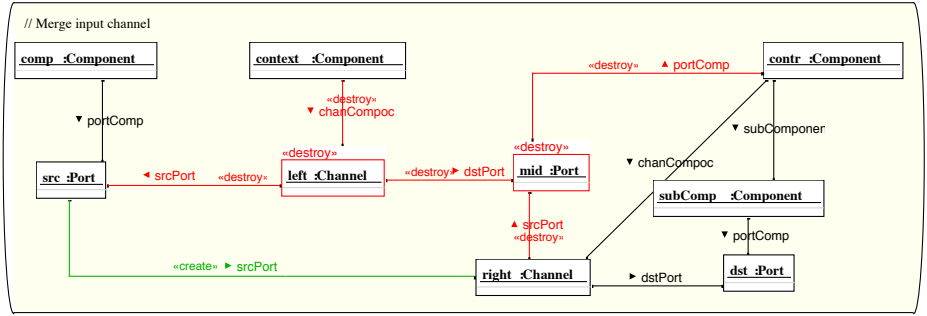

**Fig. 9.5.** Pattern-Based Specification of the Merge Rule of the Push-Down Refactoring

Since transformations basically correspond to relations between graphs, besides using these object pattern diagrams operations can be described directly in relational formalisms similar to the QVT approach [26] and its respective implementations such as ATLAS [27], F-Logics based transformation [28], or TEFKAT [29]. Furthermore, these rule-based relational approaches allow to use more declarative as well as more imperative forms of specification, e.g., providing a description of a specification in a purely declarative fashion, and an alternative, more imperative and efficient form. Strictly declarative, relational, and rule-based approaches as in [30] even allow to use a single homogenous formalism to describe the basic transformation rules and their composition. Complex analysis or transformation steps can be easily modularized since there are no side-effects or incremental changes during the transformation. Additionally, declarative approaches allow to use loose characterizations of the resulting model, supporting the exploration of a set of possible solutions to automatically search for an optimized solution, e.g., balanced component hierarchies, using guiding metrics; the set of possibile solutions can also be incrementally generated to allow the user to interactively identify and select the appropriate solution.

Technically, often a distinction between *exogenous* and *endogenous* transformations is used, depending on the characteristics of the metamodels, the source and the target of the transformation conforms to ([31], [32]). While in endogenous transformations, the source and target models are instances of the same metamodel, in the exogenous case they are instances of different metamodels. Besides the endogenous model refactorings and the exogenous model translations, model transformations are particularly helpful in the in between case of metamodel evolution. Large overlaps between source and the target domain lead to similar but differing metamodels. Here, model transformations can support the migration of models during the evolution of the metamodel, as discussed in Section 9.4.

Besides these fundamental issues of model transformation – see [31] for an overview – for the practical application also further aspects are of importance. Specifically, aspects like debugging support to trace the application of rules, analysis support to ensure syntactic and semantic correctness of transformation rules, the understandability of rules and their changeability with respect to size, complexity and degree of modularization,or the efficiency of transformations both with respect to the framework to import and export models as well as the execution of the transformation rules are gaining increasing attention.

### 9.2.3    Model Versioning and Model Merging

Model-based software engineering improves the development process by lifting the level of description from the solution domain – i.e., the domain of the implementation – to the problem domain – i.e., the domain of application – raising the level of abstraction to reduce the accidental complexity of the engineering task to focus on the essential complexity. Model analysis, e.g., conformance checking, and model synthesis, e.g., model transformation increase the degree of automation. However, raising the level of abstraction also introduces new problems.

A core aspect in the development of complex and long-lasting systems, as, e.g., generally found in embedded software systems, is the construction of those systems in incremental and often parallel steps. In traditional, code-based approaches, techniques like versioning and merging support the step-wise and distributed implementation. In a model-based development process, corresponding techniques must be supplied on model-level. While the linear structure of program code simplifies the task of comparing different fragments or merging them, the same problem of comparing or merging models is more complex due to their more general, graph-like structure.

To compute the difference between two models or to obtain a merge version of two models, the commonalities between those two models are identified via matching. To construct this matching, two different approaches are possible. If a model is described via its edit-history, consisting of the basic operations – like introducing or deleting elements or relations, changing their attributes, etc – to obtain this model, the matching essentially corresponds to identify the common operations.

However, in many cases models do not include those edit histories. Here, the matching has to be constructed by directly comparing these models; since elements of these models generally also do not maintain unique identifiers under modification – especially during deletion and re-insertion – matching has to be based on some notion of correspondence over model elements, generally based on similarities of attribute values. The most general approach to construct a matching for that case is based on a fixed-point iteration, starting with a pairwise correspondance between the elements of two models, and extending this correspondance through the relations of each model. Since this general approach is rather complex, generally heuristics are used to improve the efficiency of the matching.

The latter matching approach is, e.g., implemented in the *SiDiff* algorithm [33]. For the construction of differences on the model level, *SiDiff* has, e.g., been integrated in the MatLab environment *MATE* [7], or the UML-like FuJaBa framework [19].

## 9.3     Evolution

Model management generally handles the operations necessary to deal with models on a large scale in large projects. Equal in importance, though perhaps not equal in scale, is the need to manage models as incremental evolution is required. We discuss these issues in this section.

We use the term *evolution* here in the same sense that it is commonly used in discussions of science: incremental changes brought about as external factors change. As we discuss in this section, these external factors can include changes to the system requirements, the language used to describe requirements and models, as well as changes of style. In each of these cases, the technology used to evolve the models is largely the same. However, the techniques to evolve the models may vary.

The problem of evolution is not new to software engineering. Various approaches have been suggested to address the evolution problem in various software domains, the most prominent being schema evolution in object oriented databases. While there have been some attempts to extend these techniques to other areas such as model based software [34] [35], the nature of DSMLs and their evolution suggest that there is a need for a dedicated solution.

### 9.3.1     Evolutionary Model Development

When categorizing development processes that we use today, we find two basically different approaches with respects to models: The document / waterfall like approaches use chains of models from early requirements down to running code. When changes appear they are usually applied on the current level only (i.e., the code level) and models of earlier phases are not touched anymore. These approaches need tracking of decisions between their artifacts that then allow co-evolution of models and code with respects to new requirements etc.

The other type of software development approaches, namely the agile ones, do not use models at all. They rely on code from the beginning, and this has several advantages: Code is executable and thus provides a form of immediate feedback that non-executable models couldn't. Furthermore, tests can be written in code to and automatically rerun every single adaptation of the code. Regression tests such as these give confidence that changes to the code do not violate the requirements that are encoded in such tests.

Many users of modelling technologies desire to raise the level of abstraction at which they are programming. This requires code generators to produce software based on the semantics of the models. Assuming that such code generators provide us with the ability to "program" on the higher level using models, we can use these models to create tests as well as our system specifications. Such models can be considered "executable" (i.e., not used simply for analysis or documentation), and are therefore the principal artifacts of our construction phase.

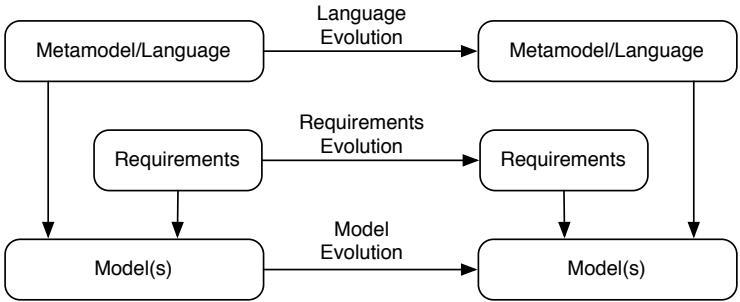

**Fig. 9.6.** There are two major preconditions for the need to evolve models. The evolution of the modelling language may invalidate structures, types, or constraints that are used and assumed by the original models. The evolution of requirements may invalidate design choices made by the original modeler. Either, or both, of these evolution types may trigger the need for model evolution.

### Elements of Model Evolution

When discussing model evolution there are a few terms that we use with a specific meaning. We discuss mainly the evolution of models that conform to a *metamodel*. This metamodel describes the language used to specify the models.

Models are frequently transformed in order to be used as an artifact later in the design chain. We use the term *model transformation* to describe rules for rewriting models. Model transformation is one technique for *model interpretation* which is a generic term that describes how meaning is given to some model.

In model transformation, we talk about transforming some *source model* ($SM$) into a *destination model* ($DM$). Rules that describe how these transformations are specified are discussed later. Destination models may conform to a different metamodel than the source models.

Now, any of these elements may change, and thus require evolutionary changes to the models. If the destination metamodel changes, then existing rules to transform source models may need to be updated. If changes are made to the source metamodel, these transformation rules may also be invalid. Similar changes may need to be made if new constraints on the structure of either the source or destination languages is anticipated.

### Actors in Model Evolution

There are several actors specific to model evolution:

– Model Designer (a.k.a., Original Modeler): This actor utilizes a modelling language to develop a model. This model is the *original* or *source* model in the discussions of this section, and represents an authoritative version of the intent of the model designer.
– Model Evolver: This actor takes the *original* model and evolves it as necessary. Reasons for this evolution may include bug fixes, changes in requirements, changes in the language, etc.
– Language Designer: This actor created the modelling language used by the original modeler (the *original modelling language*).

 – Language Evolver: This actor modified the original modelling language.
 – Requirements Specifier: This actor created the *original* requirements against
   which the original model was designed or tested.
 – Requirements Evolver: This actor modified the original requirements.

These roles may be played by the same person for small projects, but for many
large projects there will be several persons playing the various roles, or in fact
several persons playing the same role. As such, the issue of maintaining intent
throughout the design and construction phases, as well as subject to various
evolutionary paths, is of great importance. However, not all evolutionary trans-
formations need human intervention, as we discuss next.

### 9.3.2   Automating Evolutionary Transformations

Although there are many motivators for transformations, we examine here a few
of the most common reasons for transforming models, namely changes in requi-
rements, and changes in style that fall under the category of "refactoring." As
discussed earlier, the changes in requirements may result in changes of seman-
tics, while refactoring changes are (by definition) limited to behavior-preserving
transformations. We also will discuss how transformations of a domain-specific
modeling environment can be further automated, based on the strong typing in
their metamodel. Another major reason for model transformations is changes in
the domain itself, but the complexity of this issue deserves additional explana-
tion, and it is covered in Section 9.4.

   How can we automate the transformation of models, or automate the creation
of transformation specifications, to aid model evolution? In fact, the former
is achievable if the burden of creating the patterns and other assorted rules
is placed on a model transformation expert. The automatic creation of these
specifications (rules) is computationally feasible, though it brings into question
whether semantics are maintained, constraints are satisfied, etc.

### Evolution through Refactoring

In an agile development process (where only executable artifacts are created) we
must accept that the executable artifacts—in our case, the models—need to be
capable of modifications similar to those of software refactoring [36].

   Model refactoring is a form of model evolution, where the semantics of the
model remain the same, but the structure changes. We note that this defini-
tion is not always used for refactoring (specifically the clause that semantics
are preserved), but we use this definition to avoid confusion with other kinds
of model evolution where semantic preservation is not the goal. It may also be
useful for the reader to consider the notion of software refactoring, where auto-
matic changes are made to a software project based on renaming a variable, or
changing a class name (i.e., all dependencies and uses are appropriately renamed
throughout the software project). A comprehensive listing of refactoring types
for *software* is presented by Fowler et al. in [36].

Similar changes in the domain of model refactoring are possible. For example, in [37] a tool is discussed that shows how to specify model transformations that will extract a superclass from a set of selected classes on the metamodel level. This application towards domain-specific modelling languages provides an avenue to maintain the types of a modelling language, while streamlining the metamodel definition.

There are also many applications to model refactoring for domain-independent models created using UML modelling tools. In [38], the authors show how the development of new metamodels (representing the "source" patterns to be matched, and "destination" patterns to be the resultant models) can work with existing UML models.

**Evolution for Requirements Changes**

The application of agile techniques in a model-driven sense means that models are used in the beginning of development, and continued to be used throughout the project lifecycle. Since the definition of agile development is to discover requirements along the way, or refine them as they are clarified by some customer, a model-driven approach *must* be robust to changes in the requirements during the development phase [39].

This means that some models, though correct when they were built, may be subject to new requirements now, or in the future. Automating these transformations based on updated specifications changes is not feasible, and many specifications languages such as $\mathbb{Z}$ [40] are somewhat infamous for an inability to synthesize the system for which they describe requirements. This is not as much a limitation of those requirements languages, but rather a reminder that updates from changes in formal requirements should be made by a knowledgeable actor—namely the model evolver, consulting with the requirements evolver.

Consider a model-based design of a controller for an unmanned aerial vehicle [41]. The controller is designed to satisfy a certain requirement for rise-time and overshoot of the vehicle state. However, if that requirement changes, the controller design must also change. This may be simply a change in values (changing the rise time, for example), but if the change in requirements is significant, or disruptive, the design may also need to change.

In any case, one major benefit of model-based engineering is that the controller is synthesized from the model (either in software, configuration for a runtime tool, etc.). However, there is the question of *are any existing requirements unmet, after the evolution performed for requirements changes?* In order to answer these questions, the model must be subjected to regression tests that verify requirements satisfaction for the model. Previous work in regression tests for models [42] concentrates on the differences between two models, thus reducing the number of regression tests that need to be run. If models are appropriately tracked, then work such as this can dramatically reduce the time to confirm that the models still conform to the requirements.

It is important to point out, though, that without existing tests for meeting requirements, that there can be no certainty that the models *as built* even conform to the specified requirements.

**Domain-Specific Model Evolutions**

A domain-specific modeling language provides special advantages to evolving models within the same language. This is because the specific goal of domain-specific modeling is to provide a programming environment (language) that represents the domain concepts as programming primitives. This is true both for domain-specific models that are in a metaconfigurable environment such as GME [43] or AToM³ [21], but also for sophisticated user environments such as LabView and MATLAB/Simulink, who present domain-specific toolboxes for creating models.

### 9.3.3   Semantics of Evolution

A conflict of intentions comes to the forefront when evolving models. Specifically, the following question must be answered: *as these models evolve, should the original intent of the modeler be preserved, or does this evolution overrule any original intent?* Answering this question can be difficult, and in many cases requires judgement to be made by the model evolver.

There are many changes to the language, and some in the requirements, that can be automated such that no model evolver need be "in the loop" to confirm semantic interaction. However, there are many other evolutionary transformations that can be automated, but *certainly* need intervention by a modeler to confirm intent of evolution.

Consider the example of port specialization, where an object of kind `Port`, specialized into two types, `InPort` and `OutPort`, should be rewritten as either one of these types. The metamodel is shown in Figure 9.7a, and explicitly shows that a `Component` can contain zero or more *Port* objects. These `Port` objects can be associated through a `BufferedConnection`. As the metamodel shows, this connection can either be made to any to `Port` objects (who would then play the role of `src` or `dst` in that association), or explicitly between an `OutPort` (playing the role of `src`) and an `InPort` (playing the role of `dst`).

In order to rewrite models to (essentially) make `Port` an abstract type, some context is needed to determine whether existing `Port` objects are likely to be playing the role of `src` or `dst`. Certainly a patten could be written such that all models playing the role of `src` become `OutPort` models. However, this gives rise to two obvious problems: (1) what if there exists a `Port` that is not playing *any* role in an association, and (2) what if there exists a `Port` object that is playing *both* the `src` and `dst` role in two separate associations? These possibilities are shown in Figure 9.7b.

Port `p12` of `C1` exemplifies the issue of determining the type of a `Port` that plays *no* role in any association. A casual human observer may infer that the placement near another `InPort`, or its proximity to the left of `C1` would imply an `InPort`, but at this point, some human must enter the decision loop to help determine this, or a policy of "all unmatched ports become `InPort` models" must be adopted.

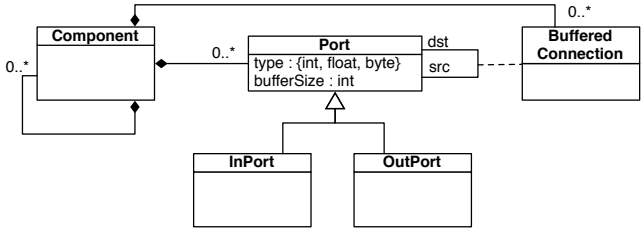

(a) The metamodel allows objects of kind `Port`, which is specialized as `InPort` and `OutPort`.

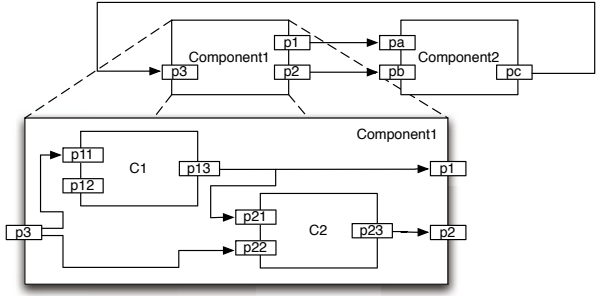

(b) A model built using the metamodel in (a). The contents of `Component1` are shown to display the additional associations in which its `Port` objects play a role. The "arrow" end of the connections represent the `dst` role.

**Fig. 9.7.** (a) A metamodel allowing port interconnection between components. (b) A model built using the metamodel in (a). The "arrow" end of the connections represent the `dst` role. If all ports in the various components are of type `Port`, then how to automatically convert them into `InPort` or `OutPort` while maintaining the original modeler's intent?

Port `p3` of `Component1` plays the role of `dst` in its connection with port `pc` of `Component2`[1], and plays the role of `src` in its connection with ports `p11` of `C1`, and `p22` of `C2`. It is clear to a casual human observer that `p3` is an `InPort`, but making this determination based on context requires careful specification of many partially ordered matches.

In [44] the issue of additional context is solved, where the containment relationships of various `Port` objects, the `Component` to which they belong, and the `Component` to which the `BufferedConnection` belongs all play a role in matching the pattern. However, as that work also states, the first problem (`Port` objects that play no role at all) still requires some actor, namely the model evolver, to make a decision on which type the `Port` should assume upon transformation.

Although there are a significant number of corner cases such as these, where the difficulty of creating an evolution strategy that preserves the original intent is called into question, there are a significant number of evolutions that can be created computationally. We will discuss the techniques useful for automating such transformations in Section 9.3.2, but now we turn to the mechanics most commonly used in performing the transformations.

---

[1] As a shorthand, we use directed connections to show `src` and `dst` roles, where the arrow resides on the `dst` role.

## 9.4    Modelling Language Evolution

Development environments evolve as tool vendors constantly improve their tools. Although programming languages have become quite stable, we cannot claim the same about modelling languages: OMG languages are still the subjects of major upgrades and the UML models created so far need to be upgraded as well. If models are used to program against libraries or components, we have another source of model upgrades, although they do not change the structure of the models, but their vocabulary. With domain-specific languages, [3], we will need even more profound evolution techniques, as DSLs are usually made for single domains/companies or even projects and thus have the tendency to strongly evolve over time.

Model evolution is the transformation of domain models that were created under a language, L, to be well-formed and conform in the successor language, L'. Of particular importance is the question of the semantics of the models under each language. Existing work in the area of domain model evolution focuses on the techniques and methods for synthesizing transformations based on changes in the metamodel. Sprinkle's thesis [44] provides an academic perspective (for the mechanics of synthesizing such transformations, see [45]). Techniques for the graphical specification of the semantics of a modeling language (i.e., the code generator associated with a metamodel) can be found in [46]. A proposal by Bell [47] advocates the creation of a catalog of grammar transformations that are capable of automating the evolution of DSL programs.

We divide these kinds of model evolution tasks into two categories: *syntactic model evolution* and *semantic model evolution.*

### 9.4.1    Syntactic Model Evolution

Syntactic model evolution is a transformation or set of transformations that rewrites a model to conform to its new metamodel. It is useful for this set of transformations to be partially ordered, to permit deterministic results of the application. We do not require syntactic model evolution to be an atomic translation, but we instead depend upon the definition of a deterministic syntactic transform set to produce a logically atomic translation (though perhaps in several phases which produce intermediate or temporary artifacts).

Syntactic model evolution *only* guarantees that the model as transformed will be syntactically valid (i.e., conform to the new metamodel). As such, a trivial solution is to delete all models in the repository, but such a solution is clearly not acceptable. However, it does present the difficulty of using syntactically driven transformations to a model evolver after the language has evolved. Consider the frustration of loading a model into a modeling environment, only to realize that one single model is causing an exception. If deleting that model allows the model evolver to load the models, they may decide that they have completed evolution, but may have violated a large set of requirements in deleting that model.

There is a concrete example, which we can draw from our previous discussion of Figure 9.7. If we interpret this issue as removing the type `Port` from the modeling language, and replacing it with two types, `InPort` and `OutPort`, we are now dealing with a model that does not conform to its evolved metamodel. Namely, the existence of objects of type `Port` violates the abstract syntax requirements.

There are two trivial solutions which satisfy the requirements for syntactic model evolution; (1) transform all models of type `Port` to `InPort`, and delete all `BufferedConnections`; (2) transform all models of type `Port` to `OutPort`, and delete all `BufferedConnections`. Of course, nontrivial solutions will yield a "correct" result, which we discuss in the next subsection.

Nonetheless, syntactic model evolution is a powerful tool for an advanced model evolver. With expert knowledge of the metamodel, and of the state of the model, syntactic model evolution can provide a rapid way to reload existing models, regardless of their semantics. One reason for this might be that changes to the language were to remove types that were no longer relevant: so, deleting those types is appropriate. Another reason might be that the models were developed in the very early stages of the project, and they will all have to be examined anyway, so any models preserved will be used, but deleting models that violate new language conditions is not disastrous, as they will be recreated with new types.

Of course, much of this depends on the size of the database of models as well. For model databases of size 10-20, a careful, complete, model evolution may take weeks to create, but the models can be rebuilt in a few hours. All of these considerations are relevant to the decision of the model evolver.

### 9.4.2   Semantic Model Evolution

Semantic model evolution is a transformation or set of transformations that rewrites a model to have the same semantics in its new language that it had in its original language. The observant reader will note that syntactic model evolution is guaranteed in a semantic model evolution process, because for preserved semantics, the evolved model must conform to the evolved language.

It is undoubtedly true that syntactic model evolution can result in a semantic model evolution, if properly applied by the model evolver. This allows standard model transformation techniques to be applied to evolving models, if the transformation patterns are appropriately designed. Such is the work by Sprinkle in [45], and by Karsai et al. in [46].

The semantic model evolution problem is also similar to the tool integration problem. In [48], Tratt motivates the benefits that model transformation offers for tool integration. The two issues of syntactic and semantic interoperability of tools is discussed in [49], which also advocates model transformation as the conversion mechanism between tool models.

**Questions of Semantics**

What happens when more than one semantic domain exists for a particular language? If multiple semantic interpretations exist, then each member of the set of semantics *must* be satisfied in order to claim that a semantic model evolution has taken place. This issue can be extremely difficult, as changes in the language may negate the ability to attach a models semantics to a particular semantic domain.

Consider the domain of hybrid systems, where transitions between states represent discrete switches in the continuous dynamics of a system. If two particular semantic domains, simulation using one tool, and analysis using another, depend on portions of the modeling space not used by the other tool, than any removal of those portions of the modeling space may affect one tool, but not the other. A concrete example is to remove the invariant set from the modeling language: analysis tools require this set in order to verify whether the system state travels outside this set, while simulation tools can still operate without that set. It is possible, therefore, to still utilize one semantic domain, but not the other, with existing models by just deleting the invariant set from all objects.

The lesson here is that the more semantic domains to which a modeling language attaches, the more difficult it is to evolve that modeling language. For domain-specific languages, the issue is both more complex, and simpler, in that by constraining the language to a particular domain, the risk of that domain changing is reduced: however, if changes do propagate to that domain, the language *must* be evolved in order to maintain its intuitive relationship with the domain types.

There are additional difficulties introduced when changes in the constraints of a language (only) may in fact cause certain models to no longer satisfy those constraints. However, this problem can easily be checked by loading the models and running the constraint checker to determine any violations. It is still an open problem to understand what changes of constraints can be directly used to transform models where violations occur, or to predict that no violations will occur.

### 9.4.3  Techniques for Automated Model Evolution

Automating model evolution in the face of language evolution is tricky, if the goal is semantic model evolution. Nonetheless, there exist techniques for helping to determine how models should evolve in order to maintain semantics across evolution.

Differences between the original and evolved metamodel can help identify elements that have changed. This does require, however, some fairly advanced algorithms for detecting changes [50], unless such changes are recorded as they are made. In this sense, the correspondence models of a triple-graph grammar may provide sufficient indication of change, but may not provide a sufficient indication of what transformations are required for a semantic model evolution.

Examining the semantic domains to which the modeling language attaches, and how that semantic domain has changed between the original and evolved

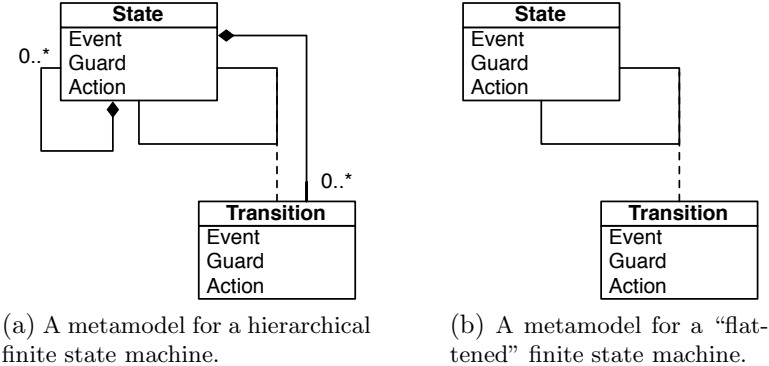

(a) A metamodel for a hierarchical finite state machine.

(b) A metamodel for a "flattened" finite state machine.

**Fig. 9.8.** (a) This metamodel allows containment of states by other states.(b) This evolved metamodel removes the ability to contain states hierarchically. Looking at the code generator, which removes checking for `State` containment, may imply that such semantics are no longer important, rather than implying that such semantics should be reapplied without hierarchy.

metamodel, is another aid. For example, if the algorithm to generate code or models has simply renamed `Type1` to `Type2`, then this may be sufficient to evolve the models (change all models of type `Type1` to be of type `Type2`).

However, there may be subtle issues even with this approach, as we show with the metamodel and evolved metamodel shown in Figure 9.8. If the algorithm to generate code removes the check for `State` objects contained within a `State`, and the metamodel indicates that containment of `State`s within each other is no longer allowed, then a naive approach could simply remove all `State` objects contained within another `State`. Unfortunately, this can also be interpreted as a requirement that an existing hierarchical state machine must be flattened. Algorithms exist that can refactor state machines [13] to be semantically equivalent, but the model evolver must realize that this is the requirement.

### 9.4.4   Step-By-Step Model Evolution

In the previous sections, we discussed modeling language evolution methods that are able to handle arbitrarily large gap between the original and the evolved language. However, in most of the practical cases, modeling language evolution does not happen as an abrupt change in a modeling language, but in small steps instead. This also holds for UML: apart from adding completely new languages to the standard, the language has been changing in rather small steps since its first release.

This assumption facilitates further automation of the model evolution by tools for metamodeled visual languages [51] [52]. The main concepts of a step-by-step evolution method is depicted in Figure 9.9.

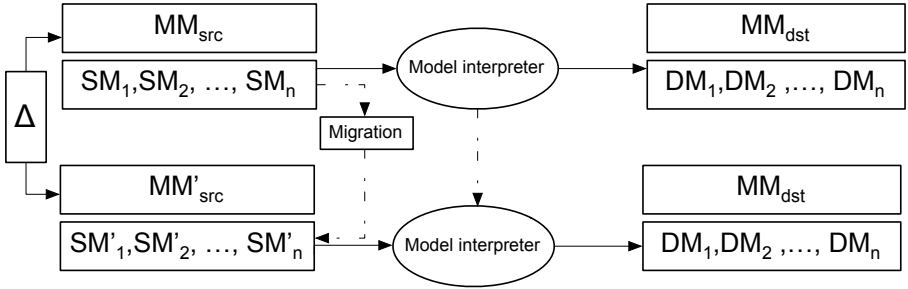

**Fig. 9.9.** Step-By-Step Evolution Concepts

The backbone of the diagram is a well-known DSL scenario depicted in the upper half of the figure. When a domain-specific environment is created, it consists of a metamodel ( $MM_{src}$), which may have arbitrary number of instance models ($SM_1$, $SM_2$, ...,$SM_n$. The models need to be processed or transformed ("interpreted"), therefore, an interpreter is built. The interpreter expects that its input models are compliant with $MM_{src}$. In parallel, the output models of the interpreter must be compliant with the target metamodel $MM_{dst}$. The inputs of the interpreter are $MM_{src}$, $MM_{dst}$, and an input model $SM_i$, and produces an output model $DM_i$.

The objective is to migrate the the existing models and interpreters to the evolved language. For the sake of simplicity, we assume that only the input modeling language evolves, de output model remains the same. The evolved counterparts are denoted by adding a prime to the original notation. In the evolution process, we create the new (evolved) metamodel ($MM'_{src}$). We assume that the changes are minor enough both in size and nature, such that they are worth being modeled and processed by a tool, rather than writing a transformation from scratch to convert the models in the old language to models in the evolved language. This is a key point in the approach.

Having created the new language by the evolved metamodel, we describe the changes in a separate migration DSL (Model Change Language, MCL). This is denoted by $\Delta$, and it represents the differences between $MM_{src}$ and $MM'_{src}$. Besides the changes, this model contains the actual mappings from the old models to the evolved ones, providing more information how to evolve the models of the old language to models of the new language. Given ($MM_{src}$), ($MM'_{src}$), and $MCL$, a tool can automatically migrate the models of the old language to models of the evolved language.

Furthermore, also based on the ($MM_{src}$), ($MM'_{src}$), and $MCL$, it is possible to migrate the model interpreter. Since it cannot be expected that the way of processing the new concepts added by the evolution can be invented without human interaction, the tool can produce an initial version of the evolved interpreter only. A usual implicit assumption here is that the language elements appearing in both the old an the evolved model should be processed in the same way. Moreover, using this assumption and the $MCL$ model, the interpreter for the parts of the old language that have been unambiguously changed by the evolution can also be generated.

In the following two sections, we present a possible realization for both the change description and the interpreter evolution.

**Describing the Changes**

Recall that our approach uses a DSL to describe the changes between the original and the evolved metamodels. This raises a a natural question: what sort of changes should be described and how? The second part of the question is partly answered: one can use a DSL to describe these changes. However, there is another criterion: the change description must hold enough information to facilitate the automated evolution of the already existing models ($SM_n$).

The first part of the question can only be answered by the practice. Below we show the construct we distilled by the experience gained in one of our research project and described in detail in [51].

Figure 9.10 outlines the structure of an MCL rule. For the sake of simplicity, we use the convention that elements on the left-hand-side of the $MapsTo$ relationship belong to the original metamodel ($MM_{src}$), and the right-hand-side elements are taken from the evolved metamodel ($MM'_{src}$). The most important concept in MCL is the $MapsTo$ connection. This connection originates from a class in the original metamodel, and points to a class in the evolved metamodel. One can assign conditions and commands written in imperative code to a mapping.

The basic operations provided by MCL are as follows:

(i) Adding elements, such as class, associations, and attributes. In Figure 9.11, we add a new element called *Thread* within a Component, along with a constraint that every *Component* must contain at least one *Thread*. The old models can then be migrated by creating a new *Thread* within each *Component*.

(ii) Deleting elements: classes, attributes, or associations. It is important that deleting elements means that we do not need that information anymore, we can lose it. If the information contained by an element is used in the evolved model, the element should not be deleted. The operation needed in this case falls into the next category. As Figure 9.12 shows, deletion is modeled with mapping a class to the null class.

(iii) Modifying elements, such as attributes and class names. The conditional mapping to new or other model elements also falls into this category. Figure 9.13 depicts a migration rule for a prevalent model refactoring case: a class becomes abstract base, and the existing instances are migrated as the instances of the new subclasses, based on certain conditions, typically, the attribute values of the instances. The conditions assigned to the $MapsTo$ connections specify which mapping must be performed. The attribute calculations and other projections from the old class to the new ones are described by the commands assigned to the connection $MapsTo$.

(iv) Local structural modifications. If the operations detailed above need to be performed in a certain context, it can be defined by the *WasMappedTo* connection. Figure 9.14 shows an example, where we short-circuit the containment hierarchy. The intent of the migration is to move all instances of *Class* up in the containment hierarchy: the instances should be contained by *ParentParent* instead of *Parent*. *WasMappedTo* does not specify an operation, it ensures that *that ParentParent* originally containing the *Parent* should be the parent of the *Parent*'s children. Recall that the left-hand-side of the figure references classes from the source metamodel, whereas the other side references classes in the evolved metamodel, thus, the name conflict does not matter in this case.

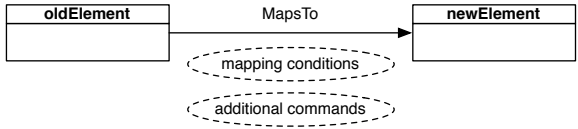

**Fig. 9.10.** Schematic description of an MCL rule

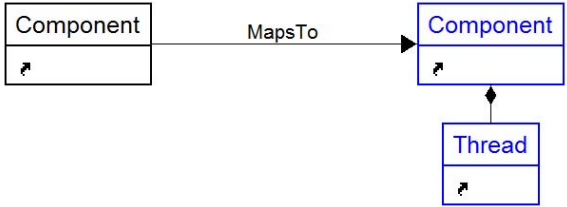

**Fig. 9.11.** Addition rule

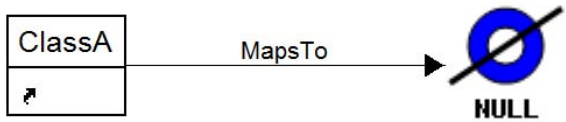

**Fig. 9.12.** Deletion rule

Given a the old metamodel, the evolved metamodel and the MCL description, a code generator is able to create executable code that migrated the $(SM_n)$ models to the new DSL defined by the evolved metamodel.

**Evolving the Model Transformations**
As it is shown in Figure 9.10, not only the models, but the model interpreter must also evolve. The Universal Model Migrator Interpreter Evolver (UM-MIE) package is a tool to semi-automate the interpreter migration process. The tool takes the old metamodel $(MM_{src})$, the evolved metamodel $(MM'_{src})$, the $\Delta$ model in MCL, the destination metamodel $(MM_{dst})$ and the old model transformation rules under the assumption that the destination model does not evolve ($MM_{dst}$ is identical to $MM'_{dst}$).

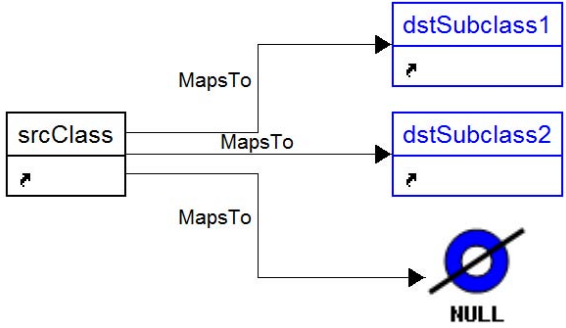

**Fig. 9.13.** Modification rule

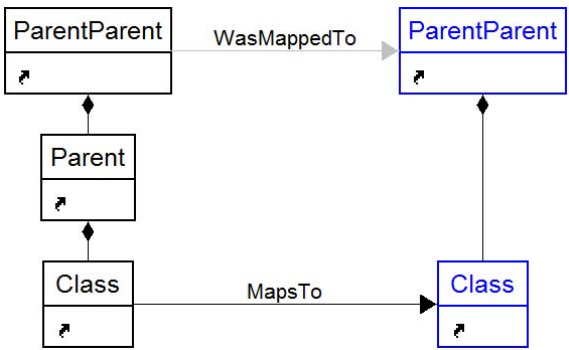

**Fig. 9.14.** Rule with context

We assume that the rule nodes in the transformation reference the input and output metamodel classes. The tool traverses the transformation rules, and takes each rule node to process the referenced metamodel classes. If these classes are in the destination model ($MM_{dst}$), or they were not changed by the MCL model, they remain intact. If a class has been deleted, the reference in the rule is set to null reference. Moreover, a warning is emitted that the null reference in the rule must be resolved manually. If a class has changed unambiguously by modification, such as renaming attributes, the tool automatically updates the rules. If there are multiple mappings such as in Figure 9.13, the tool emits a warning that the mapping should be done manually. Since the tool traverses the old transformation rules, the additions are not handled by the tool, their evolution must be performed by hand.

The UMMIE tool performs all the changes that must always be made. There are cases, in which there are several options, it depends on the intentions of the transformation developer. The main future direction of the tool is to provide "design patterns" for these cases, exposing the options to the developer, and after the selection, the evolution step is completed automatically.

## Acknowledgements

The work presented is partially sponsored by DARPA, under its Disruptive Manufacturing Program. The views and conclusions presented are those of the authors and should not be interpreted as representing official policies or endorsements of DARPA or the US government.